\documentclass[aps,twocolumn,eqsecnum,floatfix,showpacs,superscriptaddress]{revtex4}

\usepackage{graphicx}
\usepackage{amsmath}
\usepackage{amsfonts} 
\usepackage{bm}

\begin{document}

\title{Transmission probability through a L\'{e}vy glass and comparison with a L\'{e}vy walk}
\author{C. W. Groth}
\affiliation{Instituut-Lorentz, Universiteit Leiden, P.O. Box 9506, 2300 RA Leiden, The Netherlands}
\affiliation{SPSMS, UMR-E 9001, CEA-INAC/UJF-Grenoble 1, 38054 Grenoble, France}
\author{A. R. Akhmerov}
\affiliation{Instituut-Lorentz, Universiteit Leiden, P.O. Box 9506, 2300 RA Leiden, The Netherlands}
\author{C. W. J. Beenakker}
\affiliation{Instituut-Lorentz, Universiteit Leiden, P.O. Box 9506, 2300 RA Leiden, The Netherlands}
\date{May 2011}
\begin{abstract}
Recent experiments on the propagation of light over a distance $L$ through a random packing of spheres with a power law distribution of radii (a socalled L\'{e}vy glass) have found that the transmission probability $T\propto 1/L^{\gamma}$ scales superdiffusively ($\gamma<1$). The data has been interpreted in terms of a L\'{e}vy walk. We present computer simulations to demonstrate that diffusive scaling ($\gamma\approx 1$) can coexist with a divergent second moment of the step size distribution ($p(s)\propto 1/s^{1+\alpha}$ with $\alpha<2$). This finding is in accord with analytical predictions for the effect of step size correlations, but deviates from what one would expect for a L\'{e}vy walk of independent steps.
\end{abstract}
\pacs{05.40.Fb, 05.60.Cd, 42.25.Bs, 42.25.Dd}
\maketitle

\section{Introduction}
\label{intro}

A random walk with a step size distribution that has a divergent second moment is called a L\'{e}vy walk \cite{Shl95a,Met00,note3}. A L\'{e}vy \textit{glass} is a random medium where the separation between two scattering events has a divergent second moment. The term was coined by  Barthelemy, Bertolotti, and Wiersma \cite{Bar08}, for a random packing of polydisperse glass spheres. They measured the fraction $T$ of the light intensity transmitted through such a random medium in a slab of thickness $L$, and found a power law scaling $T\propto 1/L^{\gamma}$ with a superdiffusive exponent $\gamma\approx 0.5$ --- intermediate between the values for ballistic motion ($\gamma=0$) and regular diffusion ($\gamma=1$).

The simplest theoretical description of propagation through a L\'{e}vy glass neglects correlations between subsequent scattering events. The ray optics of the problem is then described by a L\'{e}vy walk, with a power law step size distribution $p(s)\propto 1/s^{1+\alpha}$, $0<\alpha<2$. The experiment \cite{Bar08} was interpreted in these terms, with $\alpha=1$ and $\gamma=\alpha/2$ the expected transmission exponent.

Correlations between scattering events in a L\'{e}vy glass dominate the dynamics in one dimension \cite{Bee09,Bur10}. Although correlations were expected to become less significant with increasing dimensionality \cite{Kut98,Sch02}, Buonsante, Burioni, and Vezzani  \cite{Buo11} have calculated that the transmission exponent $\gamma$ should remain much larger than would follow from a L\'{e}vy walk with uncorrelated steps. In particular, a saturation at the diffusive value $\gamma=1$ for $\alpha>1$ is predicted --- even though the second moment of the step size distribution becomes finite only for $\alpha>2$.

To test these analytical predictions for the effect of correlations, we have simulated the transmission of classical particles through a L\'{e}vy glass, confined to a slab of thickness $L$. Both a two-dimensional (2D) system of discs is considered and a three-dimensional (3D) system of spheres. We find a power law scaling $T(L)\propto 1/L^{\gamma}$ with an exponent $\gamma$ that lies well above the $\gamma=\alpha/2$ line expected for a L\'{e}vy walk. In particular, we obtain a saturation of $\gamma$ at the diffusive value of unity well before the $\alpha=2$ threshold is reached of a divergent second moment. 

The outline of the paper is as follows. Since our aim is to compare the L\'{e}vy glass simulations with the predictions for a L\'{e}vy walk, we need analytical results for uncorrelated step sizes. These are summarized in the Appendix and referred to in the main text. We start off in Sec.\ \ref{Levyglass} with a description of the way in which we construct and simulate a L\'{e}vy glass on a computer. The results presented in that section are for 2D, where the largest systems can be studied. We turn to the 3D case in Sec.\ \ref{Levyglass3D} and compare with the experiments \cite{Bar08}. We conclude in Sec.\ \ref{sec_discuss}. 

\section{L\'{e}vy glass versus L\'{e}vy walk}
\label{Levyglass}

\subsection{Construction}
\label{construct}

A L\'{e}vy glass \cite{Bar08,Bar10} is a random packing of transparent spheres with a power law distribution of radii,
\begin{equation}
n(r)\propto 1/r^{1+\beta}.\label{nrdef} 
\end{equation}
Light propagates without scattering (ballistically) through the spheres and diffusively (mean free path $l_{\rm mfp}$) in the region between the spheres. The probability to enter a $d$-dimensional sphere of radius between $r$ and $r+dr$ is proportional to the fraction $n(r)dr$ of spheres in that size range, multiplied by the area $\propto r^{d-1}$. The ballistic segments (steps) of a ray inside a sphere of radius $r$ have length $s$ of order $r$. The sphere radius distribution \eqref{nrdef} therefore corresponds to the step size  distribution \cite{note2}
\begin{equation}
p(s)\propto 1/s^{1+\alpha},\;\;{\rm with}\;\;\beta=\alpha+d-1.\label{alphabetadef}
\end{equation}

Particles propagating through a L\'{e}vy glass therefore have the same distribution of single step sizes as in a L\'{e}vy walk, but the joint distribution of multiple step sizes is different: While in a L\'{e}vy walk the steps are all uncorrelated (annealed disorder), in the L\'{e}vy glass the configuration of spheres is fixed so subsequent steps are correlated (quenched disorder). 

\begin{figure}[tb]
\centerline{\includegraphics[width=0.6\linewidth]{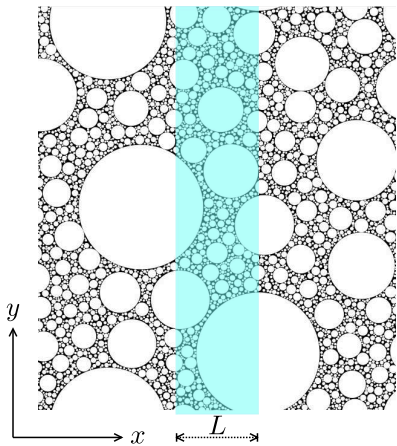}}
\caption{\label{fig_glass}
Two-dimensional L\'{e}vy glass, consisting of a random packing of discs with a power law distribution of radii ($\alpha=0.7$, $f=0.86$, $r_{\rm max}/r_{\rm min}=100$). The blue region defines a slab of thickness $L$. This is the unconstrained geometry, because the maximum disc size can be larger than $L$. 
}
\end{figure}

We discuss in some details the construction of the 2D L\'{e}vy glass, see Fig.\ \ref{fig_glass} --- the 3D version is entirely analogous. We start by generating discs of (dimensionless) radius
\begin{align}
r_{k}={}&r_{\rm max}\left[1+\frac{k}{k_{\rm max}}(r_{\rm max}^{\beta}-1)\right]^{-1/\beta},\nonumber\\
&\;\;k=0,1,2,\ldots k_{\rm max}.\label{rkdef}
\end{align}
The $k_{\rm max}+1$ discs have radii ranging from $r_{\rm min}\equiv 1$ to $r_{\rm max}\gg 1$, and in this size range their distribution follows the power law \eqref{nrdef}. The average area of a disc is
\begin{equation}
\langle A\rangle=\frac{\pi\beta}{|2-\beta|}\max(1,r_{\rm max}^{2-\beta}).\label{Abar}
\end{equation}
The entire L\'{e}vy glass occupies an area of dimension $W\times W$ in the $x-y$ plane, with periodic boundary conditions and $W$ about 10--100 times larger than $r_{\rm max}$. For a random packing we place the discs at randomly chosen positions in the order $k=0,1,2,\ldots$ (so starting from the largest disc). If disc number $k$ overlaps with any of the discs already in place, another random position is attempted. For each disc some $10^{4}$ attempted placements are made. If they are all unsuccessful, the entire construction is started over with a smaller value of $k_{\rm max}$.

\begin{figure}[tb]
\centerline{\includegraphics[width=0.8\linewidth]{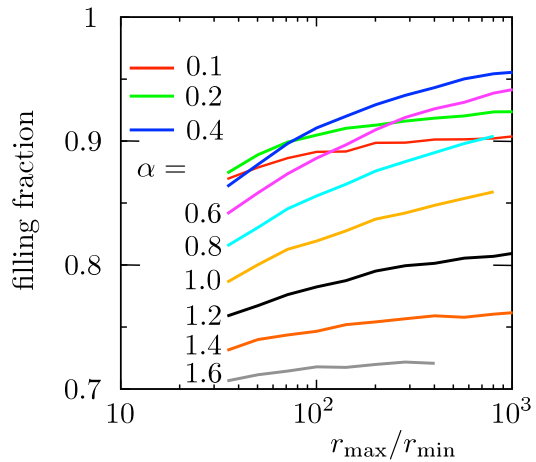}}
\caption{\label{fig_filling}
Filling fraction of the 2D L\'{e}vy glass as a function of the ratio $r_{\rm max}/r_{\rm min}$ of largest and smallest disc size, for several values of the parameter $\alpha$. 
}
\end{figure}

The density of the packing is quantified by the filling fraction
\begin{equation}
f=k_{\rm max}\langle A\rangle/W^{2}.\label{fdef}
\end{equation}
For each simulation we strove for maximal $f$, by maximizing $k_{\rm max}$. The maximal filling fraction increases with increasing ratio $r_{\rm max}/r_{\rm min}$, as illustrated in Fig.\ \ref{fig_filling}. For the smallest $\alpha$, below about $0.4$, we could not reach as dense a packing as for larger $\alpha$, basically because there are too few small discs. Somewhat larger filling fractions would be reachable by moving the discs after placement, but we did not attempt that.

\subsection{Dynamics}
\label{sec_dyn}

The ballistic dynamics inside the spheres consists of chords of varying length $s$ traversed in a time $s/v$. The diffusive dynamics in between the spheres is modeled by a Poisson process: isotropic scattering in a time interval $dt$ with probability $vdt/l_{\rm mfp}$. The mean free path $l_{\rm mfp}=r_{\rm min}/2$ is chosen such that there is, on average, one scattering event between leaving and entering a sphere. We take the same refractive index (and velocity $v$) inside and outside the spheres, so the ray is not refracted at the interface. 

\begin{figure}[tb]
\centerline{\includegraphics[width=0.9\linewidth]{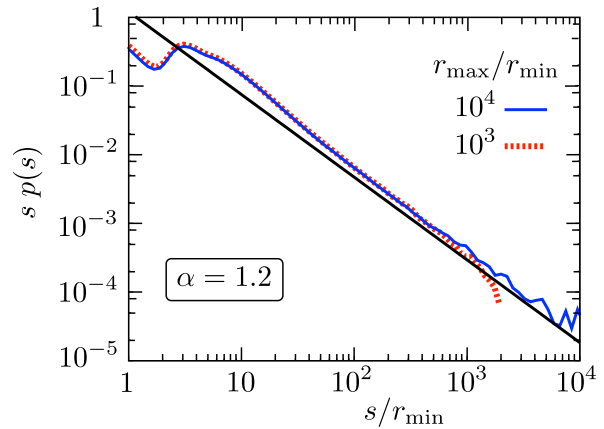}}
\caption{\label{fig_stepsize}
Step size distribution for a random packing of discs with radius distribution \eqref{nrdef} (for $\beta=2.2$, so $\alpha=1.2$). The numerical results are shown for two values of the maximum disc radius ($r_{\rm max}/r_{\rm min}=10^{4}$ and $10^{3}$, with $f=0.83$ and $0.80$, respectively). The black solid line is the expected distribution \eqref{nrdef}. 
}
\end{figure}

In Fig.\ \ref{fig_stepsize} we show the step size distribution $p(s)$ for a 2D L\'{e}vy glass with disc radius distribution \eqref{nrdef}, for $\beta=2.2$. It follows closely the L\'{e}vy distribution \eqref{alphabetadef}, with the expected parameter value $\alpha=\beta-1=1.2$ (solid line). 

We do not find the pronounced oscillations in $p(s)$ which in Ref.\ \onlinecite{Bar10} complicated the determination of $\alpha$. These oscillations appear due to coarse graining of the disc size distribution $n(r)$ and vanish if a finer distribution of disc sizes is used.

\begin{figure}[tb]
\centerline{\includegraphics[width=0.9\linewidth]{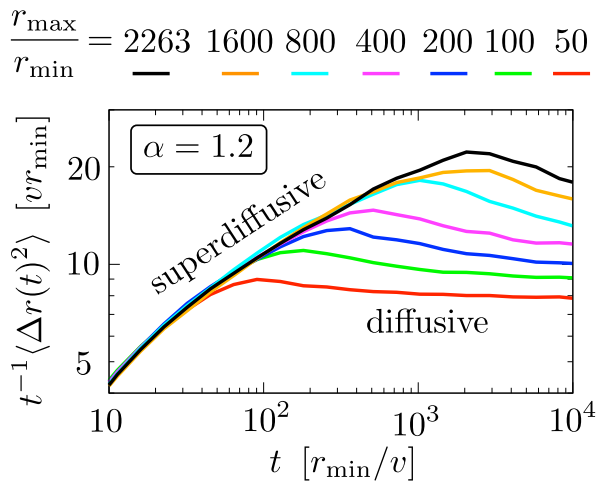}}
\caption{\label{fig_diffusion}
Time dependence of the mean square displacement (divided by $t$ so that saturation indicates diffusive scaling). The curves are the results of a numerical simulation in a 2D L\'{e}vy glass with different values of $r_{\rm max}/r_{\rm min}$, at fixed $\alpha=1.2$.
}
\end{figure}

The time dependence of the mean squared displacement $\langle \Delta r(t)^{2}\rangle$ is shown in Fig.\ \ref{fig_diffusion}, for the same $\alpha=1.2$. A particle was started at a random position $\bm{r}(0)$ in the inter-disc region, and then its position $\bm{r}(t)$ at time $t$ (either inside or outside a disc) gives the displacement $\Delta r(t)=|\bm{r}(t)-\bm{r}(0)|$. The average $\langle\cdots\rangle$ is over some $10^{4}$ initial positions. In accord with previous simulations \cite{Bar08,Bar10}, regular (Brownian) diffusion with $\langle \Delta r(t)^{2}\rangle\propto t$ is reached for times $t\gtrsim r_{\rm max}/v\equiv t_{\rm D}$, set by the time needed to traverse the largest disc. For $t<t_{\rm D}$ the mean squared displacement increases more rapidly than linearly (superdiffusion).

\begin{figure}[tb]
\centerline{\includegraphics[width=0.8\linewidth]{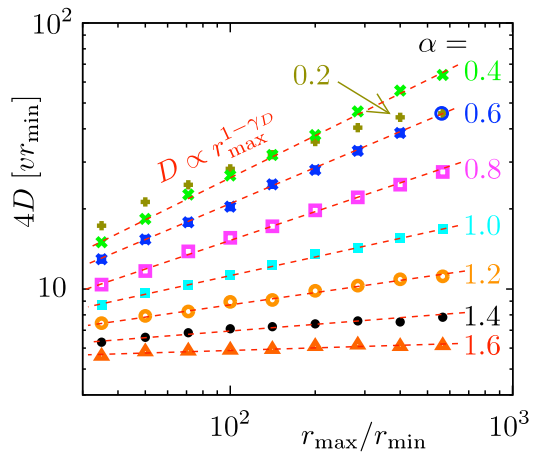}}
\caption{\label{fig_Deff}
Diffusion coefficient \eqref{Ddef} in the Brownian regime, estimated from the large-$t$ slope of the mean square displacement (corresponding to the large-$t$ saturation value in Fig.\ \ref{fig_diffusion}). Each set of colored data points represents one value of $\alpha$, with different values of $r_{\rm max}/r_{\rm min}$. The power law scaling \eqref{gammaDdef} (red dotted lines) determines the scaling exponent $\gamma_{D}$.
}
\end{figure}

The limiting slope of the mean square displacement for $t\gg t_{D}$ gives the diffusion constant in the Brownian regime,
\begin{equation}
D=\lim_{t\rightarrow\infty}\frac{1}{2dt}\langle \Delta r(t)^{2}\rangle.\label{Ddef}
\end{equation}
As shown in Fig.\ \ref{fig_Deff}, this diffusion constant has a power law dependence on $r_{\rm max}$,
\begin{equation}
D(r_{\rm max})\propto r_{\rm max}^{1-\gamma_{\rm D}},\label{gammaDdef}
\end{equation}
with $0<\gamma_{D}<1$. (For the smallest $\alpha=0.2$ no clear power law scaling was observed.)

\subsection{Transmission probability}
\label{transprobglass}

For the transmission problem we need a slab of variable thickness $L$. We distinguish two ways of constructing this geometry. One way is to obtain the slab from the entire L\'{e}vy glass by cutting out the region $0<x<L$ (blue strip in Fig.\ \ref{fig_glass}). We call this an unconstrained geometry, because $r_{\rm max}$ is not constrained to be smaller than $L$. The alternative constrained geometry (used in the experiments \cite{Bar08}) requires that the spheres all lie fully inside the slab, thereby restricting $r_{\rm max}<L/2$. We consider the transmission probabilities in the unconstrained and constrained geometries in separate subsections, both for 2D. (Results for 3D are presented in the next section.)

\subsection{Unconstrained geometry}
\label{unconstrained}

A lower limit $T_{\rm ball}$ to the transmission probability $T_{\rm uncon}$ in the unconstrained geometry follows by considering only ballistic rays, which pass through the region $0<x<L$ without a single scattering event. As explained in the Appendix, see Eq.\ \eqref{qdef}, this probability is directly related to the step size distribution,
\begin{equation}
T_{\rm ball}=\frac{1}{\langle s\rangle}\int_{L}^{\infty}dx\int_{x}^{\infty}ds\,p(s).\label{Tballdef}
\end{equation}

We take the step size distribution \eqref{alphabetadef} with an upper cutoff at $s_{\rm max}\simeq r_{\rm max}\gg L$ and a lower cutoff at $s_{\rm min}\simeq 1$. Then Eq.\ \eqref{Tballdef} evaluates to
\begin{align}
T_{\rm ball}&\simeq \frac{r_{\rm max}-\alpha^{-1}L^{1-\alpha}r_{\rm max}^{\alpha}}{r_{\rm max}-r_{\rm max}^{\alpha}}\nonumber\\
&\xrightarrow{r_{\rm max}\gg L} \left\{\begin{array}{ll}
1&{\rm for}\;\;0<\alpha<1,\\
L^{1-\alpha}&{\rm for}\;\;1<\alpha<2.
\end{array}\right.
\label{Tball}
\end{align}

Since $T_{\rm ball}\leq T_{\rm uncon}\leq 1$ we can immediately conclude that $T_{\rm uncon}=1$ for $0<\alpha<1$. For $1<\alpha<2$ the power law scaling $T_{\rm uncon}\propto 1/L^{\gamma}$ must satisfy $\gamma\leq\alpha-1$. This holds irrespective of correlations between multiple steps, since these cannot affect $T_{\rm ball}$. If we neglect these correlations, we may equate $T_{\rm uncon}$ to the transmission probability $T_{\rm eq}$ of a L\'{e}vy walk with equilibrium initial conditions (see App.\ \ref{sec_eq}).  In view of Eq.\ \eqref{Teqresult}, this leads to $\gamma=\alpha-1$. We believe this result to be quite robust, since even if correlations do play a role, it is likely that they slow down the superdiffusion \cite{Kut98,Sch02}, so they would not lead to a smaller $\gamma$.

\begin{figure}[tb]
\centerline{\includegraphics[width=0.9\linewidth]{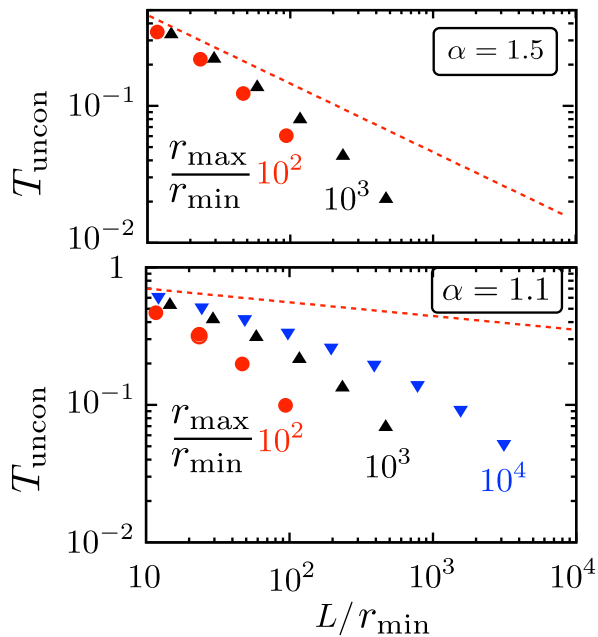}}
\caption{\label{fig_Teq2}
Transmission probability $T_{\rm uncon}$ through a 2D \textit{unconstrained} L\'{e}vy glass, for different values of the maximum disc radius $r_{\rm max}$. The dotted line is the predicted scaling $T_{\rm uncon}\propto L^{1-\alpha}$ in the $r_{\rm max}\rightarrow\infty$ limit.
}
\end{figure}

In Fig.\ \ref{fig_Teq2} we show the $L$-dependence of $T_{\rm uncon}$ for two values of $\alpha$, resulting from a numerical simulation of an unconstrained 2D L\'{e}vy glass. This is data up to $r_{\rm max}=10^{4}$ for $\alpha=1.1$ and up to $r_{\rm max}=10^{3}$ for $\alpha=1.5$, which is at the upper limit of our computational resources. As expected from the L\'{e}vy walk (Fig.\ \ref{fig_Teq}), the convergence to the $r_{\rm max}\rightarrow\infty$ limit is very slow, and we are not able to conclusively test the predicted asymptote.

\subsection{Constrained geometry}
\label{constrained}

For the construction of a constrained L\'{e}vy glass we limited the maximum disc radius to $r_{\rm max}=L/4$ and ensured that all discs fit inside the slab of thickness $L$. The corresponding random walk would be a truncated L\'{e}vy walk with maximum step size $s_{\rm max}\simeq L/2$. From the analysis in App.\ \ref{trunc_noneq} we would therefore expect a $T\propto 1/L^{\alpha/2}$ scaling of the transmission probability --- if correlations between step sizes would not matter.

\begin{figure}[tb]
\centerline{\includegraphics[width=0.9\linewidth]{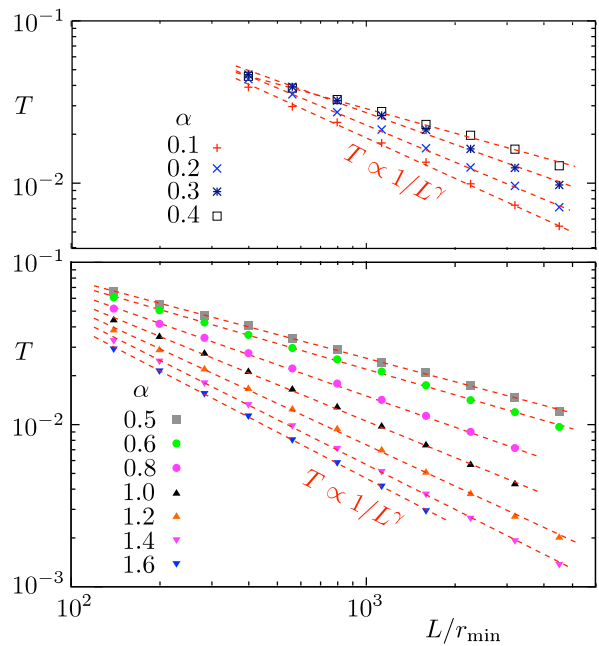}}
\caption{\label{fig_2dT}
Transmission probability through a 2D constrained L\'{e}vy glass as a function of the thickness of the slab, for different values of the step size exponent $\alpha$. The dotted lines are a linear fit to the data points, determining the transmission scaling exponent $\gamma$. ( The data is split over two panels, to avoid overlap.)
}
\end{figure}

\begin{figure}[tb]
\centerline{\includegraphics[width=0.8\linewidth]{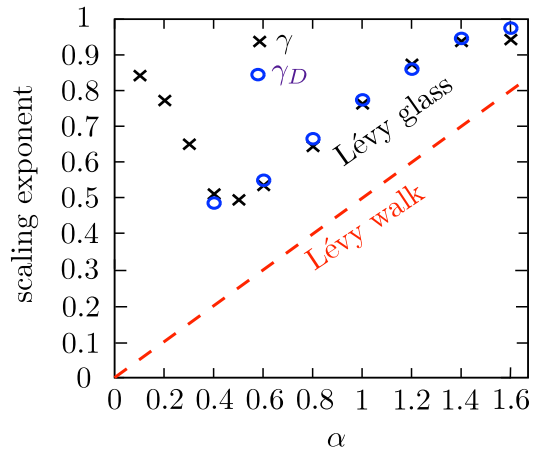}}
\caption{\label{fig_2dgamma}
Exponents $\gamma$ and $\gamma_{D}$, governing the scaling of the transmission probability \eqref{gammadef} (crosses) and diffusion constant \eqref{gammaDdef} (circles). These are the results of a simulation of a 2D constrained L\'{e}vy glass (see Figs.\ \ref{fig_Deff} and \ref{fig_2dT}). The red dashed line is the prediction \eqref{Talpha2smax} for a L\'{e}vy walk with nonequilibrium initial conditions.
}
\end{figure}

In Fig.\ \ref{fig_2dT} we show the scaling of the transmission probability,
\begin{equation}
T\propto 1/L^{\gamma},\label{gammadef}
\end{equation}
as it follows from the simulation. The power law scaling applies to somewhat less than two decades in $L$ for $\alpha\gtrsim 0.5$ (lower panel), and to one decade for smaller $\alpha$ (upper panel). In Fig.\ \ref{fig_2dgamma} we give the resulting exponent $\gamma$ as a function of $\alpha$. 

In the same figure we show the scaling of the diffusion exponent $\gamma_{D}$, from Eq.\ \eqref{gammaDdef}. (There we could only obtain a power law scaling for $\alpha\gtrsim 0.4$.) As expected from the identification of $T\simeq D(L)/L\propto 1/L^{\gamma_{D}}$, one has in good approximation
\begin{equation}
\gamma=\gamma_{D}.\label{gammagammaD}
\end{equation}

\section{Comparison with experiments}
\label{Levyglass3D}

\begin{figure}[tb]
\centerline{\includegraphics[width=0.8\linewidth]{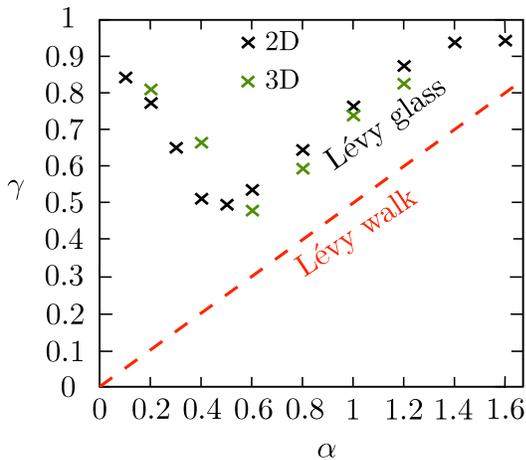}}
\caption{\label{fig_3dgamma}
Comparison of the $\alpha$-dependence of the transmission exponent $\gamma$ for a 2D and 3D L\'{e}vy glass. Both data sets lie well above the $\gamma=\alpha/2$ line of a L\'{e}vy walk.
}
\end{figure}

The numerical data shown so far was for a 2D L\'{e}vy glass of discs. We have also performed simulations for a 3D L\'{e}vy glass of spheres, in the constrained geometry with $r_{\max}=L/4$. We went up to $L/r_{\rm min}=1132$ for $\alpha\leq 0.8$ and up to $L/r_{\rm min}=800$ for $\alpha=1$ and $1.2$. (Larger values of $\alpha$ could not be simulated reliably.) Although the systems are smaller in 3D than in 2D, the results are quite similar, see the comparison in Fig.\ \ref{fig_3dgamma} of the $\alpha$-dependence of the transmission exponent $\gamma$ for a 2D and a 3D L\'{e}vy glass. In particular, for both 2D and 3D the results for $\gamma$ lie well above the $\gamma=\alpha/2$ line.

We can now compare directly with the 3D experiments \cite{Bar08}, which obtained $\gamma=0.5$ within experimental accuracy for $\alpha=1$. Our simulation, in contrast, gives for $\alpha=1$ a value for $\gamma$ which is about 50\% higher. We cannot attribute the difference to finite-size effects, since the 3D simulation reaches the same range of system sizes as the experiment. There are aspects of the experiment which are not present in the simulation (notably absorption), but we believe that the difference is mainly due to an irregularity in the experimental sphere size distribution.

\begin{figure}[tb]
\centerline{\includegraphics[width=0.9\linewidth]{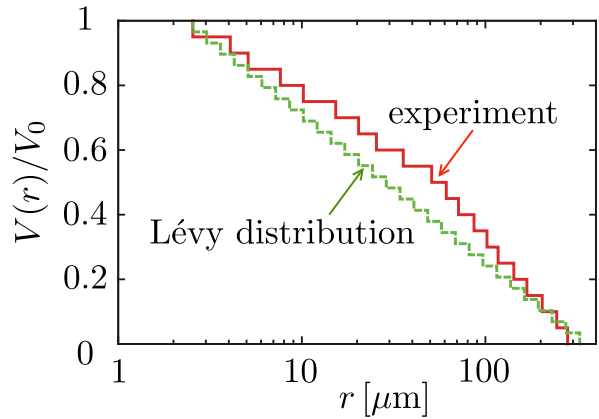}}
\caption{\label{fig_histogram}
Sphere volume distribution used in the experiment \cite{Bar08} (red solid histogram) and for an $\alpha=1$ L\'{e}vy distribution (green dashed histogram).
}
\end{figure}

To visualize the irregularity we plot in Fig.\ \ref{fig_histogram} the quantity
\begin{equation}
V(r)=\tfrac{4}{3}\pi \int_{r}^{\infty} r'^{3}n(r')\,dr',\label{Vrdef}
\end{equation}
which is the cumulative volume enclosed by spheres with radii greater than $r$. This is a decreasing function of $r$, from $V(r_{\rm min})=V_{0}$ (the total sphere volume) down to $V(r_{\rm max})=0$. For the L\'{e}vy distribution with $\alpha=1$ in 3D we have $n(r)\propto r^{-4}$, cf.\ Eqs.\ \eqref{nrdef} and \eqref{alphabetadef}, hence $V(r)$ should decrease linearly as a function of $\log r$,
\begin{equation}
V(r)=-\frac{V_{0}}{\log(r_{\rm max}/r_{\rm min})}\,\log(r/r_{\rm max}).\label{Vralpha1}
\end{equation}

As shown in Fig.\ \ref{fig_histogram}, the experimental sphere size distribution differs markedly from the expected L\'{e}vy form \eqref{Vralpha1}. Rather than a single linear dependence of $V(r)$ on $\log r$, there are two piecewise linear dependencies with a different slope, joined with a kink at $r\approx 50\,\mu{\rm m}$. This irregularity has the effect of reducing the transmission exponent $\gamma$, essentially by mimicking a system with a smaller value of $\alpha$. 

To demonstrate the effect of the kink on the transmission exponent we have simulated the experiment by constructing a random packing of spheres with the experimental size distribution (red solid histogram in Fig.\ \ref{fig_histogram}). All spheres were constrained to fit inside a slab of thickness $L$. (We took $r_{\rm max}=L/2.1$ for this simulation.) We found $\gamma=0.57$. If instead we used the proper L\'{e}vy size distribution (green dashed histogram), keeping all other parameters the same, we found $\gamma=0.72$. We believe this resolves the issue.

\section{Conclusion}
\label{sec_discuss}

In conclusion, we have found that the superdiffusive scaling $T\propto 1/L^{\gamma}$ of the transmission probability through a L\'{e}vy glass, constrained to a slab of thickness $L$, deviates substantially from what one would expect for a L\'{e}vy walk. Most significantly, the diffusive scaling ($\gamma\approx 1$) can coexist with a divergent second moment of the step size distribution ($\alpha<2$).

As a consistency check on our simulations, we have also calculated the diffusion constant $D$ from the long-time limit of the mean-square-displacement in an unbounded L\'{e}vy glass, as a function of the maximum disc size $r_{\rm max}$. We find $D(r_{\rm max})\propto r_{\rm max}^{1-\gamma_{D}}$, with $\gamma_{D}\approx\gamma$, as expected for a diffusive transmission probability $T\simeq D(L)/L$ with a scale dependent diffusion constant.

Qualitatively, our finding that diffusive scaling of $T$ can coexist with a divergent second moment of $p(s)$ is consistent with analytical calculations for $d=1$ \cite{Bee09} and $d=2,3$ \cite{Buo11}. Quantitatively, we are not in agreement: Ref.\ \cite{Buo11} finds that $\gamma$ increases monotonically for $d=2$ from $\gamma=0$ at $\alpha=0$ to $\gamma=1$ for $\alpha\geq 1$, while our simulation gives a nonmonotonic $\alpha$-dependence of $\gamma$, with a saturation for $\alpha\gtrsim 1.5$ (see Fig.\ \ref{fig_2dgamma}). The system considered in Ref.\ \cite{Buo11} is quasiperiodic (a L\'{e}vy quasicrystal), rather than the random L\'{e}vy glass studied here. Further study is needed to see whether this difference is at the origin of the different transmission scaling, or whether the difference is due to a very slow convergence to the infinite system-size limit (which we consider more likely).

\acknowledgments

This research was supported by the Dutch Science Foundation NWO/FOM.

\appendix

\section{Transmission probability of a L\'{e}vy walk}
\label{transprob}

\subsection{Formulation of the problem}
\label{formulation}

We consider a random walk along the $x$-axis with the power law step size distribution
\begin{equation}
p(s)=\frac{\alpha}{s_{0}} \left(\frac{s_{0}}{s}\right)^{1+\alpha}\theta(s-s_{0}).\label{Psdef}
\end{equation}
(The function $\theta(s-s_{0})$ equals $1$ if $s>s_{0}$ and $0$ if $s<s_{0}$.) Subsequent steps are $+s$ or $-s$ with equal probability and independently distributed. The probability density $p(s)$ decays as $1/s^{1+\alpha}$ with $\alpha>0$, starting from a minimal step size $s_{0}>0$. In between two scattering events the walker has a constant velocity of magnitude $v$. This random walk is called Brownian or diffusive for $\alpha>2$, L\'{e}vy \cite{note3} or superdiffusive for $1<\alpha<2$ and quasiballistic for $0<\alpha<1$.

The walker enters the segment $0<x<L$ by passing through $x=0$ at time $t_{i}$ and then stays in that segment until time $t_{f}$. If at $t_{f}$ it exits through $x=L$ we say the walker has been transmitted through the segment. We seek the dependence of the transmission probability $T$ on the length $L$ of the segment, for $L\gg l_{0}$. For a Brownian walk, the scaling is inverse linear: $T\propto 1/L$ if $\alpha>2$. For a L\'{e}vy walk we expect a slower power law decay, $T\propto 1/L^{\gamma}$ with $\gamma<1$. The question is how $\gamma$ varies with $\alpha<2$.

\begin{figure}[tb]
\centerline{\includegraphics[width=0.6\linewidth]{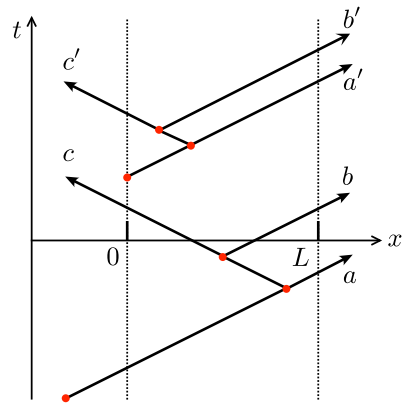}}
\caption{\label{fig_walk}
Trajectories $x(t)$ of a random walk, with scattering events indicated by red dots. All trajectories enter the segment $0< x< L$ (between dotted lines)  at $x=0$. Trajectories $a,b,a',b'$ are transmitted through $x=L$, while trajectories $c,c'$ are reflected through $x=0$. The transmission probablity $T_{\rm eq}$ averages over all trajectories (equilibrium initial conditions), while $T_{\rm noneq}$ averages only over trajectories such as $a',b',c'$ that have a scattering event upon entering the segment at $x=0$ (nonequilibrium initial conditions). 
}
\end{figure}

The answer depends on how the walker is started off initially. Following Barkai, Fleurov, and Klafter \cite{Bar00}, we distinguish equilibrium from nonequilibrium initial conditions. (See Fig.\ \ref{fig_walk}.) For equilibrium initial conditions, the walker starts off from $x=-\infty$, so that it crosses $x=0$ at some random time between two scattering events. For nonequilibrium initial conditions, the walker starts off from $x=0$ with a scattering event. We denote the transmission probabilities in these two cases by $T_{\rm eq}$ and $T_{\rm noneq}$, respectively, and consider the two cases in separate subsections. 

\subsection{Nonequilibrium initial conditions}
\label{sec_noneq}

The transmission probability $T_{\rm noneq}$ from $x=0$ to $x=L$ for a L\'{e}vy walk that starts off with a scattering event at $x=0$ has been calculated by several authors \cite{Dav97,Lar98,Bul01}. We give the most general solution of Buldyrev \textit{et al.} \cite{Bul01}. 

They assume that the walker starts with a scattering event at an arbitrary point $x_{i}$ in the segment $(0,L)$ and calculate the probability $P(x_{i})$ that the walker exits the segment through $x=L$. For $L\gg s_{0}$ and $x_{i}\gg s_{0}$ their solution \cite{Bul01} can be written in the compact form
\begin{equation}
P(x_{i})=\frac{B(x_{i}/L,\alpha/2,\alpha/2)}{B(1,\alpha/2,\alpha/2)},\label{PxiL}
\end{equation}
in terms of the incomplete beta function
\begin{equation}
B(x,a,b)=\int_{0}^{x}y^{a-1}(1-y)^{b-1}\,dy.\label{Bdef}
\end{equation}
Since $B(x,a,b)\rightarrow x^a/a$ for $x\rightarrow 0$, one arrives at the scaling $T_{\rm noneq}\propto L^{-\alpha/2}$, first obtained by Davis and Marshak from basic considerations \cite{Dav97}.

The prefactor of the power law scaling cannot be obtained directly from the solution \eqref{PxiL}, because of the limitation that $x_{i}\gg s_{0}$. For $0<\alpha<1$ we can work around this limitation by considering the first step separately. The walker starts off at $x=0$ with a step to $x_{1}>0$, chosen randomly from the distribution \eqref{Psdef} of a L\'{e}vy walk. If $x_{1}>L$ the walker is transmitted with unit probability. Otherwise, it is transmitted with probability $P(x_{1})$. 

We thus can calculate $T_{\rm noneq}$ from
\begin{equation}
T_{\rm noneq}=\int_{L}^{\infty}dx_{1}\,p(x_{1})+\int_{0}^{L}dx_{1}\,p(x_{1})P(x_{1}).\label{Tnoneq}
\end{equation}
For $\alpha<1$ the mean step size diverges, so the region $x_{1}\lesssim s_{0}$ is insignificant and we can use Eq.\ \eqref{PxiL} for $P(x_{1})$. The result is
\begin{align}
T_{\rm noneq}&=\frac{B(s_{0}/L,\alpha/2,1+\alpha/2)}{B(1,\alpha/2,1+\alpha/2)}\nonumber\\
&\xrightarrow{L\gg s_{0}}
\left(\frac{s_{0}}{L}\right)^{\alpha/2}\frac{4\Gamma(\alpha)}{\alpha\Gamma^{2}(\alpha/2)}.\label{Tnoneqresult}
\end{align}
While the exponent $\alpha/2$ holds for any $0<\alpha<2$, the prefactor is accurate only for $0<\alpha<1$. (For $\alpha>1$ we would need to know $P(x_{1})$ within the region $x_{1}\lesssim s_{0}$ in order to calculate the prefactor.)

\subsection{Equilibrium initial conditions}
\label{sec_eq}

For equilibrium initial conditions the walker crosses $x=0$ at a random time between scattering events. The first subsequent scattering event is at a point $x_{1}>0$, with probability density $q(x_{1})$. If $x_{1}>L$ the walker is transmitted with unit probability, if $0<x_{1}<L$ the transmission probability is $P(x_{1})$. Hence
\begin{equation}
T_{\rm eq}=\int_{L}^{\infty}dx_{1}\,q(x_{1})+\int_{0}^{L}dx_{1}\,q(x_{1})P(x_{1}).\label{Teqintegral}
\end{equation}

The probability density $q(x)$ is determined from the step size distribution,
\begin{equation}
q(x)=\frac{1}{\langle s\rangle}\int_{x}^{\infty}p(s)\,ds.\label{qdef}
\end{equation}
This relation between the distribution $p(s)$ of the distance $s$ between subsequent scattering events and the distribution $q(x)$ of the distance $x$ from an arbitrary point to  the next scattering event holds for any random walk with a finite average step size $\langle s\rangle=\int_{0}^{\infty}sp(s)\,ds$. For the step size distribution \eqref{Psdef} one has 
\begin{equation}
q(x)=\frac{\alpha-1}{\alpha s_{0}}\left(\frac{s_{0}}{\max(x,s_{0})}\right)^{\alpha},\;\;{\rm for}\;\;\alpha>1.\label{qresult}
\end{equation}

As emphasised in Ref.\ \onlinecite{Bar00}, the distribution $q(x)\propto 1/x^{\alpha}$ decays more slowly than the distribution $p(s)\propto 1/s^{1+\alpha}$ because the walker is more likely to cross $x=0$ during a long step than during a short step, so long steps carry more weight in $q(x)$ than they do in $p(s)$. Indeed, for $1<\alpha<2$ the first moment of $q(x)$ is infinite while the first moment of $p(s)$ is finite.

Substitution of Eqs.\ \eqref{PxiL} and \eqref{qresult} into Eq.\ \eqref{Teqintegral} gives, for $L\gg s_{0}$,
\begin{equation}
T_{\rm eq}=\left(\frac{s_{0}}{L}\right)^{\alpha-1}\frac{\pi\Gamma(\alpha)}{\alpha\sin(\alpha\pi/2)\Gamma^{2}(\alpha/2)},\;\;{\rm for}\;\;1<\alpha<2.\label{Teqresult}
\end{equation}
This scaling $T_{\rm eq}\propto 1/L^{\alpha-1}$ holds in the superdiffusive regime $1< \alpha<2$. In the quasiballistic regime the first scattering event is at $x_{1}>L$ with unit probability, 
\begin{equation}
T_{\rm eq}=1,\;\;{\rm for}\;\;0<\alpha\leq 1.\label{Teqresult2}
\end{equation}

The value $\alpha=2$ at the border between a Brownian walk and a L\'{e}vy walk requires separate consideration. While $T_{\rm noneq}\propto 1/L$ for $\alpha=2$, the transmission probability \eqref{Teqintegral} has a logarithmic enhancement,
\begin{equation}
T_{\rm eq}=\frac{s_{0}}{L}\left(1+\frac{1}{2}\ln\frac{L}{s_{0}}\right),\;\;{\rm for}\;\;\alpha=2.\label{Teqalpha2}
\end{equation}
A similar but different scaling $\propto L^{-1}\sqrt{\ln L}$ has been associated with the $\alpha=2$ L\'{e}vy walk in Ref.\ \onlinecite{Lar98}.

\subsection{Truncated L\'{e}vy walk}
\label{truncated}

A {\em truncated\/} L\'{e}vy walk has step size distribution
\begin{equation}
p_{\rm trunc}(s)=\frac{\alpha}{s_{0}} \left(\frac{s_{0}}{s}\right)^{1+\alpha}\theta(s-s_{0})\theta(s_{\rm max}-s),\label{Psdeftrunc}
\end{equation}
with a maximum step size $s_{\rm max}\gg s_{0}$. The root-mean-squared displacement $\sigma$ after a single step then has a finite value,
\begin{equation}
\sigma=\sqrt{\frac{\alpha}{2-\alpha}}\,s_{\rm max}^{1-\alpha/2}s_{0}^{\alpha/2},\label{srmsdef}
\end{equation}
much smaller than $s_{\rm max}$ for $\alpha<2$. 

The transition from a truncated L\'{e}vy walk to a Brownian walk requires $n_{\rm steps}\gg 1$ of steps, given by \cite{Man94,Shl95} 
\begin{equation}
n_{\rm steps}\simeq \frac{(2-\alpha)^{3}}{\alpha}(s_{\rm max}/s_{0})^{\alpha}.\label{nsteps}
\end{equation}
The corresponding root-mean-squared displacement $\sigma\sqrt{n_{\rm steps}}\simeq (2-\alpha)s_{\rm max}$ is of order $s_{\rm max}$ for all $\alpha<2$. We conclude that we have regular (Brownian) diffusion over a distance $L$ if $s_{\rm max}\lesssim L$. 

The transmission probability $P(x)$ for a walker starting with a scattering event at a point $x$ inside a slab of thickness $L$ (further than $s_{\rm max}$ from the  boundaries) thus follows the usual diffusive scaling,
\begin{equation}
P(x)=x/L,\;\;{\rm if}\;\;x,L-x\gtrsim s_{\rm max}.\label{Pxitruncated}
\end{equation}

\subsubsection{Equilibrium initial conditions}
\label{trunc_eq}

For equilibrium initial conditions the distribution $q(x)$ of the first scattering event follows from Eq.\ \eqref{qdef}, with $p$ replaced by $p_{\rm trunc}$. Substitution into Eq.\ \eqref{Teqintegral} then determines the transmission probability (for $L>s_{\rm max}$),
\begin{equation}
T_{\rm eq}=\int_{0}^{s_{\rm max}}dx\,q(x)P(x).\label{Teqtruncated}
\end{equation}
Eq.\ \eqref{Pxitruncated} gives $P(x)$ only for $x\gtrsim s_{\rm max}$. We will use this expression also for $x<s_{\rm max}$, and then test the approximation by comparing with numerical simulations in Sec.\ \ref{numerics}. 

If we substitute $P(x)=x/L$ we find
\begin{equation}
T_{\rm eq}=\frac{1}{2L}\,\frac{1-\alpha}{2-\alpha}\,\frac{ s_{\rm max}^{2}-s_{\rm max}^{\alpha}s_{0}^{2-\alpha}}{s_{\rm max}-s_{\rm max}^{\alpha}s_{0}^{1-\alpha}},\label{Teqtrunc0}
\end{equation}
for $0<\alpha<1$ or $1<\alpha<2$. For $\alpha=1$ or $\alpha=2$ there are logarithmic factors,
\begin{subequations}
\label{Teqalpha}
\begin{align}
&T_{\rm eq}=\frac{s_{\rm max}-s_{0}}{2L\ln(s_{\rm max}/s_{0})},\;\;{\rm for}\;\;\alpha=1,\label{Teqalphaa}\\
&T_{\rm eq}=\frac{s_{0}}{2L}\,\frac{s_{\rm max}\ln(s_{\rm max}/s_{0})}{s_{\rm max}-s_{0}},\;\;{\rm for}\;\;\alpha=2.\label{Teqalphab}
\end{align}
\end{subequations}

For fixed $s_{\rm max}$ the diffusive $1/L$ scaling holds. An anomalous scaling appears if the maximum step size $s_{\rm max}=cL$ is a fixed fraction $c<1$ of the slab thickness. Then the transmission probability through the slab depends on $L\gg s_{0}$ as
\begin{subequations}
\label{Teqtrunc}
\begin{align}
T_{\rm eq}&=\tfrac{1}{2}c^{2-\alpha}\left(\frac{s_{0}}{L}\right)^{\alpha-1}\frac{\alpha-1}{2-\alpha},\;\;{\rm for}\;\;1<\alpha<2,\label{Teqtrunca}\\
T_{\rm eq}&=\tfrac{1}{2}c\frac{1-\alpha}{2-\alpha},\;\;{\rm for}\;\;\alpha<1,\label{Teqtruncb}\\
T_{\rm eq}&=\frac{c}{2\ln(cL/s_{0})},\;\;{\rm for}\;\;\alpha=1,\label{Teqtruncc}\\
T_{\rm eq}&=\frac{s_{0}\ln(cL/s_{0})}{2L},\;\;{\rm for}\;\;\alpha=2.\label{Teqtruncd}
\end{align}
\end{subequations}
Hence $T_{\rm eq}\propto 1/L^{\max(0,\alpha-1)}$ (with logarithmic corrections for $\alpha=1$ and $\alpha=2$). This is the same scaling as for the L\'{e}vy walk without truncation (see Sec.\ \ref{sec_eq}).

\subsubsection{Nonequilibrium initial conditions}
\label{trunc_noneq}

For nonequilibrium initial conditions the transition to the regular diffusive regime happens while the walker is inside the slab. We may therefore assume that the usual diffusive scaling $T_{\rm noneq}\simeq \sigma/L$ applies (with $\sigma$ playing the role of the mean free path). In view of Eq.\ \eqref{srmsdef}, an anomalous scaling appears if $s_{\rm max}=cL$ scales proportionally to $L$,
\begin{equation}
T_{\rm noneq}\simeq (cL)^{1-\alpha/2}s_{0}^{\alpha/2}L^{-1}\propto L^{-\alpha/2}.\label{Talpha2smax}
\end{equation}
The anomalous $L^{-\alpha/2}$ scaling of Sec.\ \ref{sec_noneq} now appears as a consequence of regular diffusion with a scale dependent mean free path.

\subsection{Numerical test}
\label{numerics}

\begin{figure}[tb]
\centerline{\includegraphics[width=0.9\linewidth]{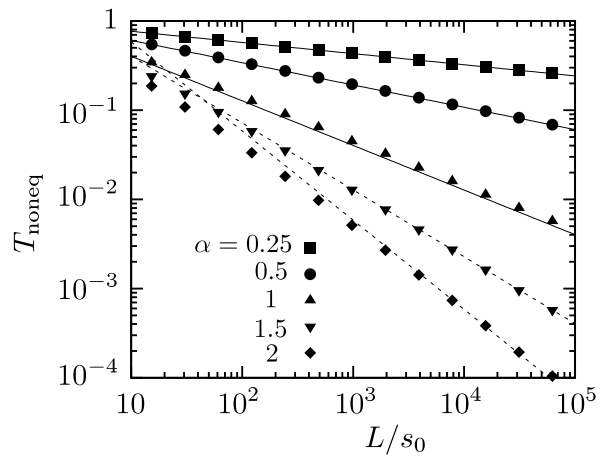}}
\caption{\label{fig_Tnoneq}
Transmission probability $T_{\rm noneq}$ of a L\'{e}vy walk through a slab of thickness $L$, for {\em nonequilibrium} initial conditions. The data points are the results of a numerical simulation, for different values of the step size exponent $\alpha$ (and fixed $s_{\rm max}\gg L$). The lines indicate the expected $L^{-\alpha/2}$ scaling. For $\alpha<1$ we also have an analytical prediction \eqref{Tnoneqresult} for the prefactor (solid lines), while for $\alpha>1$ only the exponent is known analytically so the prefactor has been fitted to the data (dotted lines).
}
\end{figure}

\begin{figure}[tb]
\centerline{\includegraphics[width=0.9\linewidth]{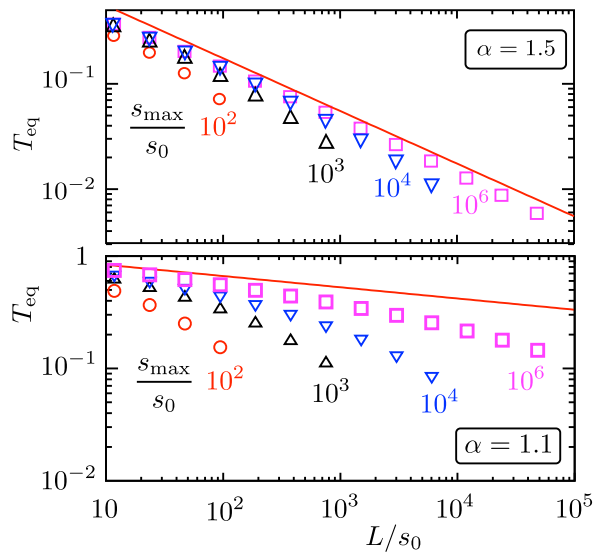}}
\caption{\label{fig_Teq}
Transmission probability $T_{\rm eq}$ of a L\'{e}vy walk through a slab of thickness $L$, for {\em equilibrium} initial conditions. The two panels are for different values of $\alpha$. The data points result from a numerical simulation, with different values $s_{\rm max}$ of the maximum step size. The solid line is the asymptote \eqref{Teqresult} for $s_{\rm max}\rightarrow\infty$.
}
\end{figure}

We have tested the analytical expressions \eqref{Tnoneqresult} and \eqref{Teqresult} by numerical simulation. Results for $T_{\rm noneq}$ are shown in Fig.\ \ref{fig_Tnoneq}. This is the nonequilibrium initial condition, where the walker starts off at $x=0$ with a step to positive $x$. The $L^{-\alpha/2}$ scaling is reproduced for all $0<\alpha< 2$, and the prefactor \eqref{Tnoneqresult} agrees well with the simulations for $0<\alpha\leq 1$.

For the equilibrium initial condition the walker starts off at a large distance from $x=0$, crossing the boundary at a random point between two scattering events. Results of numerical simulations are shown in Fig.\ \ref{fig_Teq}. Unlike in the nonequilibrium case, the convergence to the asymptotic scaling with increasing $s_{\rm max}$ is very slow, in particular for small $\alpha$.

\begin{figure}[tb]
\centerline{\includegraphics[width=0.8\linewidth]{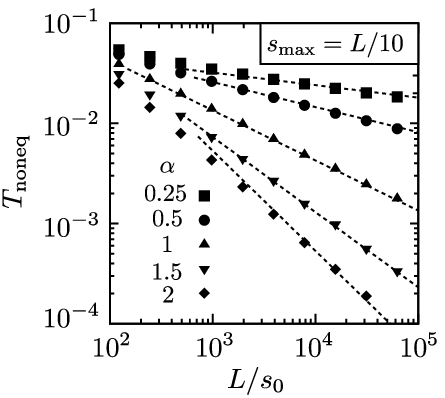}}
\centerline{\includegraphics[width=0.9\linewidth]{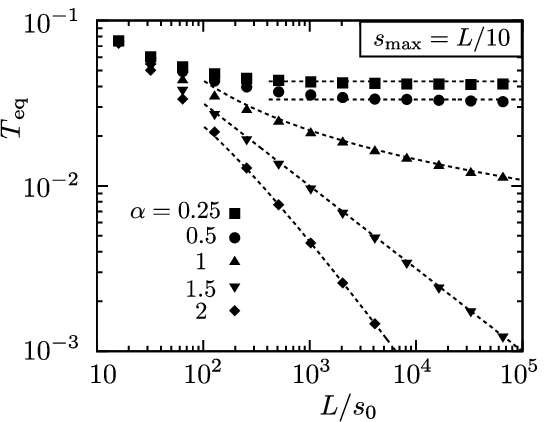}}
\caption{\label{fig_Ttrunc}
Transmission probability for a L\'{e}vy walk with maximum step size $s_{\rm max}$ that increases proportionally to $L$. The two panels (both for $s_{\rm max}=L/10$) correspond to equilibrium and nonequilibrium initial conditions. The dotted lines show the expected scaling \eqref{Teqtrunc} and \eqref{Talpha2smax}, up to a prefactor which has been fitted to the data. (For $T_{\rm eq}$ the difference with Eq.\ \eqref{Teqtrunc} is a factor of two, independent of $\alpha$.)
}
\end{figure}

We have also tested the scaling \eqref{Teqtrunc} and \eqref{Talpha2smax} for a truncated L\'{e}vy walk with a maximum step size $s_{\rm max}$ that is a fixed fraction of $L$. Results are shown in Fig.\ \ref{fig_Ttrunc} for both equilibrium and nonequilibrium initial conditions. The anomalous scaling now appears even though the diffusion is regular on the scale of $L$, because of the scale dependence of the mean free path. For both types of initial conditions the numerics follows closely the analytically predicted power laws, including the logarithmic factors for $\alpha=1,2$ in the equilibrium case. (The constant prefactors are not given reliably by the analytics.)


\begin{thebibliography}{99}
\bibitem{note3} The difference between a L\'{e}vy walk and a L\'{e}vy flight is that in the walk the steps have a duration proportional to their length, while in the flight the steps are assumed to occur instantaneously. For the transmission probability the difference does not matter, but for the mean square displacement it does.
\end{thebibliography}
\end{document}